\documentclass[usenatbib,letterpaper]{mn2e}
\usepackage{epsfig,journals}
\newcommand{\nat}{{\em Nature}}
\newcommand{\Msun}{M_\odot}
\newcommand{\ltsim}{\protect\raisebox{-0.5ex}{$\:\stackrel{\textstyle <}{\sim}\:$}}
\newcommand{\gtsim}{\protect\raisebox{-0.5ex}{$\:\stackrel{\textstyle >}{\sim}\:$}}

\begin{document}

\title[Massive Protostellar Disk Fragmentation]{Fragmentation of
  Massive Protostellar Disks} 
 
\author[Kratter \& Matzner]{Kaitlin M.~Kratter$^a$ and Christopher
 D.~Matzner$^{a}$ \\ $^a$Department of Astronomy \& Astrophysics\\ 
 \\ University of Toronto, 50 St. George Street, Toronto, ON M5S 3H4, Canada}
\date{printed \today} \maketitle

\begin{abstract}
We examine whether massive-star accretion disks are likely to fragment
due to self-gravity.  Rapid accretion and high angular momentum push
these disks toward fragmentation, whereas viscous heating and the high
protostellar luminosity stabilize them.  We find that for a broad range of
protostar masses and for reasonable accretion times, massive disks
larger than $\sim 150$ AU are prone to fragmentation.  We develop an
analytical estimate for the angular momentum of accreted material,
extending the analysis of Matzner and Levin to account for
strongly turbulent initial conditions.  In a core-collapse model, we
predict that disks are marginally prone to fragmentation around stars
of about four to 15 $\Msun$ -- even if we adopt conservative
estimates of the disks' radii and tendency to fragment.  More
massive stars are progressively more likely to fragment, and there is
a sharp drop in the stability of disk accretion at the very high
accretion rates expected above 110 solar masses.  Fragmentation may
starve accretion in massive stars, especially above this limit, and is
likely to create swarms of small, coplanar companions.
\end{abstract}

%\keywords{stars: formation -- accretion,--accretion-discs --instabilities-- stars: early-type}

\section{Introduction}\label{intro}
Advances in submillimeter telescopy have enabled the discovery of
flattened structures, in some cases clearly Keplerian disks,
surrounding massive ($\gtsim 10\Msun$) protostars
\citep{Ces2005,Ch2004,Pat2005,Beuth}.  The logical inference -- that
high mass star formation (HMSF) proceeds through disk accretion --
raises a question: can such disks process rapid mass 
accretion, or do they fragment to produce secondary stars?  We seek to
answer this question by estimating the criterion for disk
fragmentation in the vicinity of a massive star. 

If massive-star disks typically do not fragment, then disk
accretion poses little barrier to massive star formation.  Conversely
if disk fragmentation occurs, then accretion onto the central star
may be (partially) choked off, as suggested by
\cite{2005MNRAS.362..983T} in the context of low-luminosity active galactic
nuclei.   Moreover each massive star that forms from a fragmenting
disk may be surrounded by smaller stars that began as disk
fragments.  It is therefore important to evaluate models for HMSF in
light of the disk fragmentation criterion.  

The current work uses and extends the results of \citeauthor{ML2005}
(\citeyear{ML2005}, hereafter ML05), who showed that protostellar disks
around low mass stars are strongly stabilized against fragmentation by
a combination of viscous heating and irradiation by the central
protostar.  Massive star formation is fundamentally different,
however, in three important ways.  The rapid rise of luminosity
with mass implies that stellar irradiation is far more intense; this
tends to stabilize disks against fragmentation.  However, massive
stars must accrete quickly to form at all \citep[e.g.,][]{WC1987}, and rapid
accretion favors fragmentation.   These effects compete to set the
critical radius outside of which disks fragment.  Whether
fragmentation actually occurs depends on the initial disk radius, which
itself depends on the physical state of the gas prior to accretion. 
We discuss disk fragmentation in \S \ref{Dfrag}, considering the
stabilizing effect of viscous heating (\S \ref{visc}) before
incorporating irradiation by the central star (\S \ref{irrad}).  This
combination allows us to identify (\S \ref{nomodel}) the disk radius
at which fragmentation sets in.  
%% We compare this threshold to our
%% expectations of disk radii in the core-collapse and Bondi-Hoyle
%% accretion scenarios in \S \ref{core_vs_comp}.
The \cite{MT2003} core collapse model is examined in more detail in \S
\ref{coremod}: we compute angular momentum scales in \S \ref{angmom}
using formulae derived in the Appendix. In \S \ref{coreval} we
calculate expected fragmentation radii for a range of masses in the
core collapse model.

Turning to the consequences of fragmentation, we examine in \S
\ref{fragments} the likely properties of stars born within fragments
and the possibility that fragmentation limits accretion.  In \S
\ref{observations} we compare our results with observed regions of
HMSF.

\section{Disk Fragmentation}\label{Dfrag}

\subsection{Criterion for Fragmentation} \label{criteria}
We shall concentrate on fragmentation due to local gravitational
instabilities, which set in when Toomre's parameter 
\begin{equation} \label{toomQ}
Q=\frac{c_{\rm ad}\Omega}{\pi G \Sigma}
\end{equation}
descends toward unity.  Here $\Sigma$ is the disk's surface density,
$\Omega$ its orbital frequency, and $c_{\rm ad}$ is
its adiabatic sound speed.  (We shall frequently refer to the isothermal
sound speed $c_s = \gamma^{-1/2} c_{\rm ad}$ where $\gamma$ is the
ratio of specific heats.) 

Observational inferences of massive circumstellar tori \citep{Ces2005}
indicate that they may be subject to {\em global} instabilities as
well.  We discuss this possibility briefly in \S \ref{finitemass};
however the local instabilities tend to occur first, and their
relation to fragmentation is better understood.

Several authors have identified the fragmentation boundary in terms of
a cooling time.  Following \cite{Gam2001} and ML05, we convert this
criterion to a critical mass accretion rate at a given midplane
temperature.  The cooling time $\tau_c$ is the ratio of the internal
energy per area, $U=c_{\rm ad}^2\Sigma/[\gamma(\gamma-1)]$, to the dissipation
rate per unit area -- which, in steady accretion, is
\begin{equation}\label{vflux} 
2 F_v ={3\dot{M}\Omega^2\over4\pi} 
\end{equation}
where $F_v$ is the flux through each disk face. Eliminating $\Sigma$,
the maximum accretion rate is
\begin{equation} \label{MdotC}
\dot{M}_{\rm max} = {4 \gamma^{1/2} \over 3(\gamma-1) } {c_s^3\over Q \Omega \tau_c G}. 
\end{equation}
For later convenience we write this in terms of the isothermal sound
speed $c_s = \gamma^{-1/2} c_{\rm ad}$.  Extrapolating from
\cite{Gam2001}'s two-dimensional simulations for a stiff equation of
state, ML05 estimate $\dot{M}_{\rm max} = 0.89 c_s^3/G$ for the case
of a three-dimensional, $\gamma=5/3$ disk.  Using smoothed-particle
hydrodynamics, \cite{2005MNRAS.364L..56R} have simulated just such a
disk, finding $\Omega \tau_c$ to lie between 6 and 7 when it
fragments.  Assuming also $Q=1$, this implies $\dot{M}_{\rm max} =
(0.37$ to $0.43) c_s^3/G$.  For a fixed $\dot M$, the critical
temperature is then significantly (1.6 times) higher than ML05
estimated.
 
At face value this weakens the conclusion, reached by ML05, that
fragmentation is unattainable in low-mass protostellar disks.
However, more recent simulations show that disks remain stable at
shorter $\tau_c$ depending on how abruptly cooling is implemented
(E.~Harper-Clark 2006, private communication). Due to uncertainties in
the aforementioned cooling factor, and in light of the stringent
resolution requirements for collapse outlined by
\cite{2006astro.ph..9493N}, we consider the value of $\Omega \tau_c$
obtained by \cite{2005MNRAS.364L..56R} to be uncertain by up to
a factor of two. We shall therefore be conservative, by adopting the
stricter ML05 criterion that fragmentation occurs when
\begin{equation} \label{cscrit}
c_s < c_{s,\rm crit} = 1.04 (G \dot{M}_{\star d})^{1/3}. 
\end{equation}
Since the mass accretion rate is comparable to $\varepsilon c_{\rm
eff}(\mbox{core})^3/G$ in the core collapse scenario, where $c_{\rm
eff}(\mbox{core})^2$ is the ratio of pressure to density in the core,
equation (\ref{cscrit}) implies, qualitatively, that a disk fragments
if its sound speeds falls below the effective sound speed of its
parent core (see equation [\ref{cs_versus_sigma_for_frag}] below).
This point was made for thermally supported cores by ML05, and
equation (\ref{cscrit}) simply extends the rule to turbulent cores.

Equation (\ref{cscrit}) is conservative in the sense that
fragmentation may also occur at somewhat higher values of $c_s$.  It
is even more conservative given that $\gamma$ declines from $5/3$ at
the higher temperatures relevant to massive star formation.  We shall
make several other conservative estimates in order to show that disk
instability is all but inevitable during massive star formation.

When assessing disk stability in a given scenario, we first calculate
the midplane temperature profile $c_s(r)$ of the disk given its
central mass $M_\star$, central luminosity $L_\star$, and accretion
rate $\dot{M}_{\star d}$.  
For this we adopt the \citeauthor{SS1973} $\alpha$ parametrization
of viscosity, in which the steady mass accretion rate is 
\begin{equation}\label{Mdotvisc}
\dot{M}_{\rm visc} =  \frac{3\pi\alpha\Sigma c_s^2}\Omega. 
\end{equation} 
For the choice of critical temperature made in equation (\ref{cscrit})
this corresponds to $\alpha = 0.30 Q$ at the onset of fragmentation.
We keep $\alpha$ fixed at 0.30 throughout our analysis, as this
correctly reproduces the fragmentation boundary, although this is an
overestimate for stable disks. 

Having identified the {\em fragmentation radius} as the location where
the disk sound speed falls to the critical value in equation
(\ref{cscrit}), we then compare this to the characteristic disk radius
\begin{equation}\label{diskang}
R_d  = \frac{j^2}{G M_\star}
\end{equation}
for accreting gas with specific angular momentum $j$.  

We set the viscous accretion rate equal to the accretion rate from the
envelope, and let $Q=1$. We then check whether the disk can remain in
this steady state with two models for heat generation. In \S
\ref{visc} we ignore the luminosity from the protostar and find a
minimum value for $T_d(r)$ using only the viscous generation of heat.
Next, in \S \ref{irrad}, we include the flux from the protostar
received at the disk surface, and again solve for the midplane temperature as a
function of radius.

\begin{table} 
\begin{tabular}{|l|l|}
\hline
Subscript&Meaning\\
\hline
\hline
cl ....&Clump \\
$c$.....&Core \\
$\star$.....&Star\\
$d$.....&Disk \\
$\star d$....&Star-disk system\\
$f$....&Final value \\
crit..&Critical value for fragmentation\\
irr...& Stellar irradiation\\
$v$.....&Viscous flux \\
\hline
\end{tabular}
%\caption{Definitions of subscripts.}
\label{subscripts}
\end{table}

\subsection{Viscous Heating}\label{visc}
In a thermal steady state, the flux of viscous energy radiated by each
face of the disk is given by equation (\ref{vflux}).  All of the disks
considered in this paper are optically thick to their own thermal
radiation; therefore, the flux can also be derived from radiation
transfer across an optical depth $\kappa\Sigma/2$ from the disk
midplane to its surface:
\begin{equation}\label{radflux}
F_r=\frac{8}{3\tau_R} \sigma T_d^4,
\end{equation}
where $\tau_R=\kappa_R \Sigma_d/2$ is the optical depth corresponding
to the Rosseland mean opacity $\kappa_R(T_d)$. The factor $8/3$ in
equation (\ref{radflux}) is derived by assuming 
that the dissipation rate per unit mass is a constant
\citep{1997ApJ...477..398C}.  We obtain temperature dependent
opacities from \citet{2003A&A...410..611S}.  These opacities are very
insensitive to density; we adopt values for $10^{-12.5}\, \rm
g\,cm^{-3}$, an appropriate value for a disk with $Q=1$ and a period
of a few centuries.

Neglecting irradiation of the disk surface, in a steady state
$F_v=F_r$ and $\dot{M}_{\rm visc} = \dot{M}_{\star d}$.  We solve for
$\Sigma$ in equation (\ref{Mdotvisc}), and then $c_s$ from equations 
(\ref{vflux}) and (\ref{radflux}) using $\mu c_s^2 =
k_B T$ (for molecular weight $\mu$), yielding
\begin{equation} 
c_s^{10} = \frac{3k_B^4 \kappa_R \dot{M}_{\star d}^2\Omega^3}{128 \pi^2\sigma
  \mu^4\alpha }. 
\end{equation} 
This is an implicit formula for $c_s$, as $\kappa_R$ depends on
temperature.  Equating $c_s$ to $c_{s,\rm crit}$ (equation
\ref{cscrit}) implies fragmentation when
\begin{equation} \label{OmegaCriterion}
\Omega < 8.54 \left( \frac{\sigma \mu^4 \alpha G^{10/3}\dot{M}_{\star d}^{4/3}
}{k_B^4 \kappa_R} \right)^{1/3}. 
\end{equation}
In practice we calculate the run of $c_s(r)$ and $\kappa_R(r)$ 
self-consistently in order to evaluate equation
(\ref{OmegaCriterion}).   

Whereas ML05 found a unique value of $\Omega_{\rm crit}$ for optically
thick accretion disks around low-mass protostars, we shall show below
that equation (\ref{OmegaCriterion}) gives a roughly constant
fragmentation radius of about 130 AU.  The key difference is the
higher accretion rate, which implies much higher critical temperature
(hundreds of K) in massive star formation compared to $\sim
16$ K for the low mass case.  As the opacity
law does not obey $\kappa_R(T)\propto T^2$ for these higher
temperatures, the ML05 result does not hold for these more massive
stars.

\subsection{Stellar Irradiation}\label{irrad}

Stars of mass somewhat greater than $10\Msun$ undergo Kelvin-Helmholtz
contraction rapidly enough to settle onto the hydrogen burning main
sequence while still accreting.  As the main sequence
luminosity increases rapidly with $M_\star$, the consequences of
stellar luminosity become acute for massive stars.  One such
consequence is the heating of the disk midplane due to reprocessed
stellar radiation. In principle, this stabilizes disks to longer
periods (larger radii) than we found in equation
(\ref{OmegaCriterion}) by considering only viscous heating.  We
parametrize irradiation through the reprocessing factor $f$ defined
as the ratio between the incident flux $F_{\rm irr}$ normal to the
disk surface, and the spherical stellar flux at that radius: 
\begin{equation} \label{tmid}
F_{\rm irr} = \frac{fL}{4\pi R_d^2}. %  \sigma T_{\rm irr}^4
\end{equation}
A calculation of $f$ requires a model for the reprocessing of
starlight onto the disk.

\subsubsection{Infall geometry} 

ML05 model $f$ in the context of an infall geometry similar to that
used by \cite{1993ApJ...402..605W} and \cite{1993ApJ...414..773K} for
scattered-light images of protostars.  Starlight is absorbed at the
inner edge of a rotating infall envelope
\citep{1983ApJ...268..753C,1984ApJ...286..529T} whose innermost
streamlines have been removed by the ram pressure of a magnetically
collimated protostellar wind.  This removal implies a suppression of
the star-disk accretion rate relative to the rate at which mass would
otherwise accrete -- from the surrounding core, for instance, in a
core-accretion model:
\begin{equation} 
\dot{M}_{\star d} = \varepsilon \dot{M}_c; 
\end{equation} 
where $\varepsilon = \cos \theta_0$ if streamlines are removed from
all angles within $\theta_0$ of the axis. (We assume the prestellar
matter is isotropically distributed before it falls in.)

The balance of forces that determines $\varepsilon$ is modeled in
detail by \cite{MM2000} for low-mass star formation, and we expect
that their analysis holds into the massive star regime.  The location
of the innermost streamline is a simple function of $\varepsilon$: it
strikes the disk at a radius
\begin{equation} \label{Rinner}
(1-\varepsilon^2) R_d. 
\end{equation} 

The calculation of $f$ is particularly simple if the dust envelope is
(1) optically thick to stellar photons; (2) optically thin to
dust thermal radiation; and (3) not hot enough to sublimate.  Under
these conditions, ML05 find 
\begin{equation}\label{f}
f \simeq 0.1\varepsilon^{-0.35}
\end{equation}
for a reasonable range of $\varepsilon$. 

\subsubsection{Envelope self-opacity and dust sublimation}
\label{IrradProblems} 
It is important to examine the assumptions that led to equation
(\ref{f}), especially in the context of very rapid accretion onto very
massive stars.  

In \S \ref{angmom} we will estimate the typical infall column density
$\Sigma_{{\rm sph},f }$ in a core collapse scenario, and find it to be
a few times lower than the core column $\Sigma_c$, i.e., $\Sigma_{{\rm
sph},f }\simeq 0.3$ g cm$^{-2}$.  Given that the
\cite{2003A&A...410..611S} Planck opacity peaks between 3 and
16\,cm$^2$\,g$^{-1}$ for $45<T<930$ K, we expect the disk midplane to be
shielded by moderate optical depths ($\sim 1-5$) from the reprocessing
surface.  A solution to the radiation diffusion problem is beyond the
scope of this paper.  Instead we will apply equation (\ref{f}); this
is conservative, in the sense that it overestimates the stabilizing
effect of irradiation. 

A further potential complication arises for especially high accretion rates,
those in excess of $1.7\times10^{-3}\Msun$\,yr$^{-1}$.  The critical
temperature of such a disk is quite hot, $T_{\rm crit} > 1050$ K
according to equation (\ref{cscrit}).  As we shall see in \S
\ref{coreval}, the stabilizing effect of irradiation is accentuated in
these disks by a sharp drop in Rosseland opacity.  If the disk is to
reach 1050 K, however, dust in the nearby infall envelope must
approach the silicate sublimation temperature of about 1500 K.  This
leads to the disappearance of dust within the infall envelope inside a
certain {\em sublimation radius} $R_s$.  \cite{2002ApJ...579..694M}
determine $R_s\simeq 35 (L_\star/10^6L_\odot)^{1/2}$ AU by optical
interferometry (and note that this value requires the existence of
large silicate grains).  We expect that disks with $R_d>R_s$ are
relatively unaffected by dust sublimation.  We do not attempt
to calculate disk irradiation in cases where this is not true. 

\subsubsection{Incorporation into fragmentation calculation}
\label{IrradInCalc} 
Solving equation (\ref{tmid}),  we can find the equilibrium temperature
of the outer reaches of an optically thick disk for which irradiation
dominates over viscous heating.  While low-mass protostellar disks can
be stabilized in this regime (ML05), massive-star disks are not.  We
therefore account self-consistently for $T_d(R_d)$, in two steps.
The disk's effective (surface) temperature ($T_s$, say) is determined
by the requirement that it emit the viscous flux in addition to
re-emitting the incident flux: 
\begin{equation}\label{T_s}
\sigma T_s^4 = F_v + F_{\rm irr}. 
\end{equation} 
The midplane temperature $T_d$ is derived from $T_s$ from radiation
diffusion of the viscous flux across optical depth $\tau_R$, 
\begin{eqnarray} \label{combtemp} 
\sigma T_d(R_d)^4 &=& \sigma T_s^4 + \frac83 \tau_R F_v = \left(\frac83\tau_R
+ 1\right)F_v + F_{\rm irr} \nonumber \\ 
 &\simeq& \frac83 \tau_R F_v + F_{\rm irr}. 
\end{eqnarray} 
We account for the temperature dependence of the opacity when solving
this equation numerically.

As before, we identify the critical disk radius $R_{\rm crit}$ at
which $T_d(R_d) = T_{\rm crit}$.  Fragmentation occurs if the disk
extends beyond $R_{\rm crit}$.

\subsubsection{Other Considerations} \label{otherheat}
We pause to address two minor concerns: 

\noindent\emph{Shock Heating}-- Infalling matter is decelerated in an accretion
shock upon reaching the disk, and heat radiated by this shock
warms the disk surface.  However the gravitational potential at
$R_d$ is very small compared to that at the stellar surface, as $R_d\gg
R_\star$.  Moreover, the star's emission is dominated by hydrogen
burning rather than accretion, and a fair fraction of the starlight is
reradiated onto the disk surface in our model (eq.~\ref{f}).  Shock
heating is thus wholly negligible (by about four orders of magnitude,
for a $30\,\Msun$ star).

\noindent\emph{Radiation Pressure}-- When $Q=1$ the ratio of gas to radiation
pressure at a characteristic fragmentation radius is 
\begin{equation}\label{pressures}
\frac{P_g}{P_{\rm{rad}}}=\frac{3 k_B \Omega^2}{2\pi G a\mu T_{\rm
    crit}^3} = 10^{1.8} {30\,M_\odot\over M_\star} 
    \left(t_{\rm acc}\over 10^5\,\rm yr  \right)^2 \left(150
    \,\rm AU\over R_d\right)^3, 
\end{equation}
where $t_{\rm acc}$ is the duration of accretion (see \S\S
\ref{nomodel} and \ref{core-collapse}).  Radiation pressure remains
negligible out to periods of 3300 years (scaling as
$(M_\star\Sigma)^{-3/4}$ in the core model of \S \ref{coremod}).
Moreover the photon diffusion time across the scale height $H$,
$t_{\rm diff} \simeq 3 \tau_R H/c$, is a few hundred times shorter
than the orbital period.  Consequently photon pressure is irrelevant
for disk fragmentation during massive star formation.

\subsection{ Typical Parameters for Fragmentation} \label{nomodel}

Before treating the core accretion model in detail in \S
\ref{coremod}, we wish to draw a few conclusions that are reasonably
independent of a scenario for massive star formation.  We adopt in our
irradiation model a fiducial efficiency parameter $\varepsilon= 0.5$,
but we consider other values in \S \ref{CoreEffic}.

\begin{figure} 
\includegraphics[scale=.50]{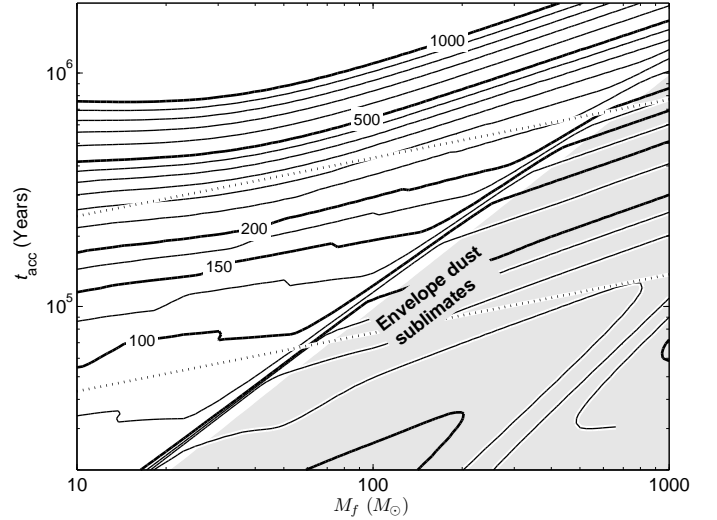}
\caption{The fragmentation radius (in AU) is shown for a range of
central star masses and accretion times. Also shown is the star
formation time in the core model ({\em dotted lines}, for
$0.3<\Sigma_{\rm cl}<3$ g cm$^{-2}$) and the region affected by dust
sublimation in the infall envelope (filled region), where our model
does not hold. The sharp kink in the lines of constant radius is due
to a drop in opacity for disk temperatures above about
1050K. }\label{riso}
\end{figure}

\begin{figure}
\includegraphics[scale=.5]{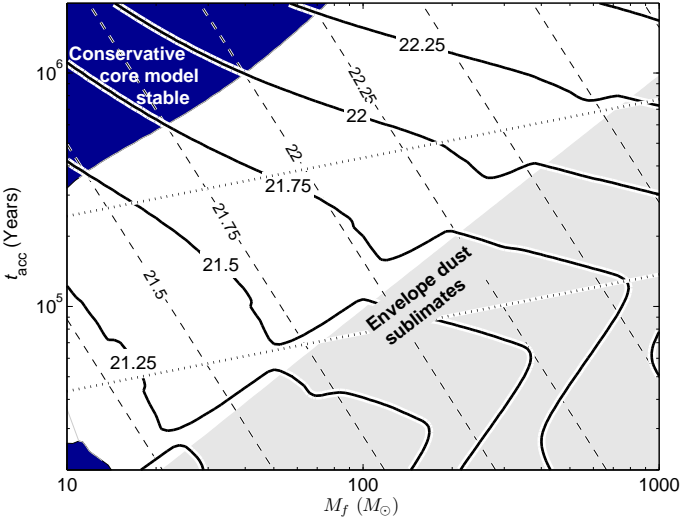}
\caption{Angular momentum and disk fragmentation.  The critical value
  of angular momentum, labeled as $\log_{10}(j_{\rm crit})$, is
  plotted as {\em bold curves} for a range of stellar masses and
  formation times, $t_{\rm acc} = 2 M_\star/(\dot{M}_{\star d})$.  {\em
  Dashed lines} show our predicted angular momentum for the
  \citet{MT2003} core model ($\varepsilon=0.5, k_\rho=1.5$). As
  discussed in \S \ref{angmom}, this is also an approximate upper
  limit to $j$ given $M_\star$ and $t_{\rm acc}$.  Disks tend to
  fragment except in the upper, dark filled region.  In the lower,
  light filled region, envelope dust grains sublimate within $R_d$;
  our model is not secure here. {\em Dotted lines} delimit core
  model formation times for $0.3<\Sigma_{\rm cl}<3$ g cm$^{-2}$.}
  \label{critang} 
\end{figure}

We begin by mapping the maximum disk radius and maximum
disk angular momentum as functions of the stellar mass
$M_\star$ and the accretion time,  %\begin{equation} 
$t_{\rm acc} = 2M_\star/{\dot{M}_{\star d}}$. 
%\end{equation} 
(The factor of two derives from the core accretion scenario of
\citealt{MT2003}; accretion time is simply a convenient
parametrization for accretion rate.)  The
results for $R_{\rm crit}$ and $j_{\rm crit}$ are shown as the solid
curves in figures \ref{riso} and \ref{critang}, respectively.  

Not all of this parameter space is relevant, however.  Observations of
protostellar outflows emerging from sites of massive star formation
imply dynamical ages of order $10^5$ years.  On both of these
figures, we highlight within the dotted lines a plausible range of
$t_{\rm acc}$ as the range of values predicted by \cite{MT2003} for core
column densities ($\Sigma_c$) of 0.3 -- 3 g cm$^{-2}$.  One may
further restrict one's attention to masses between 10 and 120 $\Msun$,
as more massive stars are not known to exist.  From this we infer:
{\em for plausible values of the accretion time scale, massive-star
accretion disks fragment for radii above about 100-200 AU, with 150 AU
being a typical threshold.}

Table (\ref{obsj})  lists our estimates for the
angular momentum of several observed disks, most of which appear
likely to fragment.

\begin{table}
\begin{tabular}{|c|c|l|}
\hline
\# in fig. \ref{frag} & $\log_{10} j$ (cm$^2$ s$^{-1}$) & Reference\\
\hline
1 & 22.3 &\cite{2005AA...434.1039C}\\
2 & 21.4 &\cite{Pat2005}\\
5 & 22.1 &\cite{2002ApJ...566..982Z}\\
6 & 22.0 & \cite{1999AA...350..197B}\\
9 & 21.6 &\cite{2005ApJ...628L.151D}\\
\hline
\end{tabular}
\caption{Estimates of angular momentum of observed disks in
figure \ref{frag} and corresponding references. Angular momentum
estimates are computed from the observed velocity gradient over the extent
of the disk. Estimates are only made for data points that showed a clear
velocity gradient associated with a disk or torus.} \label{obsj}
\end{table}

\section{Core Accretion}\label{coremod}

With these relatively model-independent results in hand, let us
evaluate disk fragmentation in the turbulent core model
\citep{1992ApJ...396..631M,MT2003}, which posits that massive stars
accrete from hydrostatic structures (cores, subscript ``$c$'') that
have assembled from an overdense region (clump, subscript ``cl'')
within a larger molecular cloud.  This is similar to the standard
model of low-mass star formation, except that turbulent motions are
subsonic within low-mass cores (e.g., ML05) and supersonic in massive
cores.

To calculate core and core collapse properties, we adopt the
\cite{MT2003} models and fiducial parameters.  Turbulent cores are
modeled as singular polytropic spheres with density profiles
$\rho(r)\propto r^{-k_\rho}$, with $k_\rho\simeq 1.5$.   Since they
are assumed to be hydrostatic \citep[see also][]{TanKrum06}, they must
be pressure confined within their parent clumps.  \citealt{MT2003}
evaluate the mean hydrostatic pressure within clumps to be 
\begin{equation} \label{P-sigma-mt03}
\bar P_{\rm cl} = 0.88 G \Sigma_{\rm cl}^2 
\end{equation} 
Further, they assume that massive cores are segregated to the centers
of clumps where the pressure is roughly twice $\bar P_{\rm cl}$.
Fixing cores' surface pressure to this value implies, for fiducial
parameters, that $\Sigma_c = 1.22\Sigma_{\rm cl}$.  
Observations imply a column density $\Sigma_{\rm cl}$ of around 1 g
cm$^{-2}$ for the clumps that host massive star formation
\citep{1997ApJ...476..730P, 1998ApJ...492..540H, 1999ApJ...525..750F,
2000ApJ...545..301K,1991ApJ...375..594V, 2001AAS...199.1404G,
1997ApJ...479L..27D, 2000ApJ...532L.109T,2004A&A...426...97F}.

As cores are bound and argued not to fragment \citep{2006ApJ...641L..45K}, their
gas either accretes on the star (a fraction $\varepsilon$) or is blown
out by the protostellar wind ($1-\varepsilon$).  The final stellar
mass is therefore $M_{\star   f} = \varepsilon M_c$, with 
$\varepsilon\simeq 0.5$, and the core radius is 
\begin{equation} \label{rcoremass}
R_c=\frac{0.040}{\varepsilon^{1/2} \Sigma_{\rm cl,cgs}^{1/2} }
\left(\frac{M_{\star f}}{30\Msun}\right)^{1/2} {\rm pc} 
\end{equation} 
where $\Sigma_{\rm cl,cgs}$ is in g cm$^{-2}$. 

In the following sections we shall refer to $\sigma=\sigma(r)$, the local
velocity dispersion of core gas at radius $r$; in a singular
polytropic core, 
\begin{equation}\label{sigma^2-SPS}
\sigma(r)^2 = {4\pi\over 6\phi_B (k_\rho-1)} {G M_c(r)\over r}
\end{equation} 
where $M_c(r)$ is the enclosed mass, and $\phi_B \equiv P/(\rho
\sigma^2)$ accounts for magnetic contributions to the total pressure;
\cite{MT2003} estimate $\phi_B\simeq 2.8$. 

We note, in passing, that the fiducial core model implicitly assumes
that stellar accretion halts due to a limited mass supply, rather than
due to the onset of vigorous stellar feedback, fragmentation of the
core or disk, or any other dynamical effect.  Alternatively one could
either have $\varepsilon=M_{\star f}/M_c\ll 1$, or one could define
the core to be the subregion that successfully accretes (for which
$\varepsilon\sim 0.5$).  In the latter choice, the
pressure-equilibrium column, $\sim 1.22 \Sigma_{\rm cl}$, would
provide only a lower limit to $\Sigma_c$.  As it is widely held
that the upper mass cutoff derives from stellar feedback, we expect
columns in excess of this lower limit to prevail for very massive
stars.  Nevertheless, we evaluate the above equations for $\Sigma_c =
1.22$\,g\,cm$^{-2}$ in the fiducial case.

\subsection{Collapse} \label{core-collapse}
\cite{MT2003} show that the accretion time is close
to the free-fall time evaluated at the core's surface density.  Their
equations (3), (4), (5), (35), and (36) specify
\begin{equation} \label{Mdot-BondiHoyle}
\dot{M}_{\star d} = \varepsilon \phi_{\rm acc} \frac{\sigma^3}{G}
\end{equation} 
where 
\begin{eqnarray}\label{phi_acc}
\phi_{\rm acc} &=& 0.71 (3.38-k_\rho)(k_\rho-1)^{3/2}(3-k_\rho)^{1/2}
\nonumber \\ &~& ~~\times \left\{ \begin{array}{cc}
  \frac{\phi_B^{3/2}}{(1+H_0)^{1/2}}, & {\rm
    magnetic} \\ 1, & {\rm nonmag.}   \end{array} \right. \nonumber \\ 
&\rightarrow&  1.9 
\end{eqnarray}
where the magnetic term includes levitation by the static field (the
factor $(1+H_0)\simeq 2$) as well as turbulent magnetic pressure (the
factor $\phi_B$).  The second line evaluates the first for a
magnetized core with $k_\rho=1.5$, for which the final accretion time
is
\begin{equation}\label{acctime}
 t_{\rm acc}=1.3\times10^5 \Sigma_{\rm cl,cgs}^{-3/4}
 \left( \frac{0.5}{\varepsilon}
\frac{M_{\star f}}{30\Msun}\right)^{1/4}\mbox 
 {  yr } 
\end{equation}
\citep{MT2003}.  Equation (\ref{phi_acc}) implies fragmentation when 
\begin{equation}\label{cs_versus_sigma_for_frag}
c_s < 1.04 (\varepsilon \phi_{\rm acc})^{1/3} \sigma, 
\end{equation}
i.e., for $c_s < 1.02\sigma$ when $\varepsilon=0.5$ and $\phi_{\rm
  acc}=1.9$.

During collapse, distribution of mass, $M_c(r)\propto
r^{3-k_\rho}$, and free-fall time, $t_{\rm ff}(r)\propto
\rho(r)^{-1/2}$, imply 
\begin{eqnarray}\label{steadymdot}
M_{\star} &=& \left({t}/{t_{\star f}}\right)^{\eta_m} M_{\star
    f}~~~~{\rm and}\nonumber \\ 
\dot{M}_{\star d}&=&\frac{\eta_m M_{\star }}{t} \propto
    M_\star^{1-1/\eta_m}  ~~~ {\rm  where} \nonumber \\ 
\eta_m&=&\frac{6}{k_\rho} -2~~ \rightarrow 2. 
\end{eqnarray} 

The accretion time is longer than the Kelvin-Helmholtz time for stars
$\gtsim 10\Msun$ \citep{WC1987}, implying that massive protostars
reach the zero age main-sequence (ZAMS) during accretion.  When
calculating the time evolution of fragmentation in \S \ref{time}, we
employ the \cite{MT2003} models for the luminosity of a massive
accreting protostar; to calculate irradiation at the end of
accretion we also include ZAMS luminosity, using formulae from \cite{Tout}. 

\subsection{Angular Momentum} \label{angmom}
To evaluate our fragmentation criterion, we require the angular
momentum of material within a collapsing turbulent core.  Although
this is not included in the \cite{MT2003} models, we shall estimate
angular momentum within these models by generalizing a calculation by
ML05 (itself an analytical version of the
\citealt{2000ApJ...543..822B} calculation, in the vein of
\citealt{1987ApJ...315..259F}).  This calculation (presented in the
Appendix) notes that a core model specifies the turbulent velocity
dispersion $\sigma(r)$ as well as the density profile $\rho(r)$.  For
a special class of velocity fields one can compute the
ensemble-averaged specific angular momentum and velocity
dispersion. The velocity field must be isotropic; it must have a
velocity difference between any two points that scales as
$\sigma(|\mathbf r_1-\mathbf r_2|)\propto |\mathbf r_1-\mathbf
r_2|^\beta $ regardless of the underlying density distribution; and
its Cartesian components must be uncorrelated (i.e., transport no
average shear stress).  The ensemble-averaged specific angular
momentum and velocity dispersion are calculated in equations
(\ref{<j^2>})-(\ref{<sigma^2>-angular}).  We define the {\em spin
parameter} $\theta_j$ for a turbulent region of size $R$:
\begin{equation} \label{theta_j_evaluation} 
\theta_j \equiv \frac{j}{R \sigma(R)} = f_j \frac{\left<j^2\right>^{1/2}}
    {R\left<\sigma(R)^2\right>^{1/2}}. 
\end{equation} 
The rightmost expression involves root-mean-square ensemble averages
over the turbulent velocity field, which are calculated in the
Appendix.  The factor $f_j$ accounts for the difference between the
ratio of rms averages and the ratio of amplitudes; since the
amplitudes are random variables, this includes both an overall offset
and a disperson.  ML05 estimate $\log_{10} f_j =
-0.088^{+0.16}_{-0.49}$ on the basis of a Gaussian model for the
velocity field.

Our evaluation of the spin parameter is presented in the Appendix and
summarized in figure \ref{jForCores} and table \ref{jOnRSigma}. For a
turbulence supported region with $\rho\propto r^{-k_\rho}$ one must
have $\sigma(r)^2 \propto G M(r)/r\propto r^{2-k_\rho}$, so
$\beta=1-k_\rho/2\rightarrow 1/4$ (for $k_\rho\rightarrow 3/2$).  In
the \citeauthor{MT2003} core collapse model, each shell of matter
accretes in sequence. For each shell, the spin parameter is 
\begin{equation}\label{thetajshell} 
\theta_j({\rm shell}) = 0.85 \beta^{0.42} f_j \rightarrow 0.47 f_j 
\end{equation} 
(see equation \ref{powerlaw-thetaj-with-beta}).  However, the last
shell to accrete naturally has the highest angular momentum; as the
disk accumulates vector angular momentum,\ref{thetajshell} may overestimate the
disk radius.  Moreover, although collapse is generally
super-Alfv\'enic, magnetic braking may sap $j$ somewhat.  A rough
lower bound on $\theta_j$ is given by the specific angular momentum of
the entire core, which corresponds to
\begin{equation}\label{thetajcore}
\theta_j ({\rm entire~core}) = 0.50\beta^{0.55} f_j \rightarrow 0.23
f_j
\end{equation}
(also equation \ref{powerlaw-thetaj-with-beta}). 

Given upper and lower limits for $\theta_j$, we must decide which
value to adopt.  The remainder of this paper is intended to establish
that fragmentation is inevitable in the outer reaches of massive-star
disks, so we shall adopt the more conservative estimate
(eq.~\ref{thetajcore}).  Please bear in mind that the
upper bound to $R_d$ is about four times larger.  The protostellar
outflow will tend to remove low-$j$ material (see also ML05), but we
expect this effect to be rather minor and do not evaluate it.

For convenience we shall define a second rotation-related quantity,
$\phi_j$, by 
\begin{equation}\label{jscale}
j = \frac{\phi_j}{\varepsilon} \frac{G M_\star}{\sigma}.
\end{equation}
The factor $\phi_j$ defined here is related to the spin parameter
$\theta_j$ by $\phi_j = R\sigma(R)^2\theta_j/[G M(R)]$.  Hydrostatic
equilibrium requires $ {R \sigma(R)^2 }/[{G M(R)}] = 1/[2(k_\rho-1)
\phi_B]$: therefore
\begin{equation} 
\phi_j = \frac{\theta_j}{2(k_\rho-1) \phi_B}  \rightarrow 0.36
  \theta_j
\end{equation}
giving $\phi_j=0.067$ in the fiducial case. 

The disk acquires a final radius which is  about 1/40th of the core
radius: 
\begin{eqnarray} \label{Rd_from_Rc}
R_{d,f}&=&\frac{ \theta_j \phi_j}{\varepsilon} R_c 
\nonumber\\
&\rightarrow&300
\left(\frac{0.5}{\varepsilon}\right)^{3/2} \left( \frac{M_{\star
f}}{30\,\Msun} \frac{1}{\Sigma_{\rm cl,cgs}}\right)^{1/2} {\rm AU}
\end{eqnarray} 
in the fiducial, conservative case given by equation
(\ref{thetajcore}).  During accretion, the disk radius remains
proportional to the current radius of accretion; therefore
\begin{equation}\label{Rd_in_accretion}
\frac{R_d}{R_{d,f}} = \left(\frac{M_\star}{M_{\star
f}}\right)^{\frac{1}{3-k_\rho}\rightarrow 2/3}, 
\end{equation}
at least on average. 

It is useful to know the column density scale in the infall for
reference in the calculation of disk irradiation.  As discussed in
ML05, this is characterized by $\Sigma_{\rm sph}(R_d)$: the column
outward from $R_d$ in a nonrotating infall of the same accretion
rate.  We find 
\begin{eqnarray}\label{Sigma_sph,f} 
\Sigma_{{\rm sph}}(R_{d,f})&=& \frac{\epsilon\phi_{\rm
  acc}}{2^{5/2}(k_\rho-1)\phi_B \theta_j} \Sigma_c 
\nonumber \\ &\rightarrow& 0.64 \Sigma_c
\end{eqnarray}
i.e., that the final infall column is comparable to the core column
(see also ML05).  Over the course of accretion,
\begin{equation}\label{Sigma_sph(m)}
\frac{\Sigma_{\rm sph}(R_d)}{\Sigma_{{\rm sph}}(R_{d,f})} = 
\left(\frac{M_\star}{M_{\star f}}\right)^{\frac{1-k_\rho}{3-k_\rho}
  \rightarrow -1/3}. 
\end{equation} 

Finally, a note on the relation between a core's density profile and
its angular momentum scale. Considering the range of values $1\leq
k_\rho\leq 2$, the angular momentum of a turbulent sphere decreases
sharply toward zero as $k_\rho$ approaches 2 and $\beta$ approaches
zero, as shown in figure \ref{jForCores}.  This trend is easily
understood: when $\beta=0$, the velocity difference between two points
is independent of their separation and must therefore be contained in
very small-scale motions as in an isothermal gas.  Indeed, the
three-dimensional power spectrum scales as $k^{-3-2\beta}$ and
contains a divergent energy at 
small scales as $\beta\rightarrow 0$.  Angular momentum is dominated
by the largest-scale motions that fit within the region of interest,
and therefore vanishes if turbulent energy appears only at small
scales.  However, a core whose hydrostatic support is effectively
isothermal ($\beta=0$) may still contain angular momentum due to {\em
background} turbulence if it is confined in a turbulent region with
$\beta>0$.  This situation holds for thermally supported
cores within turbulent molecular clouds, and formed the basis for the ML05
estimate of disk radii in low-mass star formation.   

Background turbulence may increase $j$ for turbulent cores, as well,
if the density profile flattens and the effective value of $\beta$
increases across the core boundary.  Our calculation in the Appendix
assumes that velocity field is described by the same $\beta$
everywhere, so the corrected value of $\theta_j$ should be
intermediate between the core's $\beta$ and that of the parent clump.

\subsection{Observations of Rotation in HMSF} \label{diskobs}

\cite{1993ApJ...406..528G} observe velocity gradients of dense cores,
including both low and high-mass cores, using C$^{18}$O, NH$_3$ and CS
as tracers.  They find that the ratio of rotational to gravitational
energy, which we call $\beta_{\rm rot}$ (to avoid confusion with
$\sigma\propto r^{\beta}$), takes a rather broad distribution around a
typical value of 0.02, i.e., $\log_{10}\beta_{\rm rot} = -1.7$.  In
our model for rotation within singular polytropic cores,
\begin{equation}\label{beta_rot} 
\beta_{\rm rot} = {(5-2k_\rho)^2 \over 8 (3-k_\rho) (k_\rho-1)}
     {\theta_j^2\over \phi_B}. 
\end{equation} 
Using $\theta_j=0.23 f_j$ as we estimated for the entire core
(eq.~[\ref{thetajcore}]) this gives $\log_{10}\beta_{\rm rot} \simeq
-2.1\pm 0.7$, whereas using $\theta_j = 0.47 f_j$ as appropriate to
the outer shell (eq.~[\ref{thetajshell}]), $\log_{10}\beta_{\rm rot}
\simeq -1.7\pm 0.7$.  The agreement would thus be good if the
\citeauthor{1993ApJ...406..528G} observations traced the outermost
core gas, but this seems unlikely.  A better explanation is that the
observed cores have somewhat flatter density profiles; the discrepancy
is removed if $k_\rho=1.35$ rather than 1.5 (using eqs.\ [\ref{beta}]
and [\ref{powerlaw-thetaj-with-beta}]).  As noted above, embedding the
core within a clump medium that has a flatter density profile (and
higher $\beta$) would have the same effect.  Protostellar outflows
also raise $j$ slightly by removing material on axis.  Of course, our
model for core angular momentum is based on an idealized turbulent
velocity spectrum, and undoubtedly involves some error.

We include observations of disks and toroids in massive star forming
regions in figure (\ref{frag}) and table \ref{reftab}. However it
should be noted that such comparisons are limited due to uncertainty
in both the values of quoted parameters and in the actual phenomena
being observed.  Only recently has it been possible to achieve the
resolution and sensitivity required to constrain models of massive
star formation.  Many uncertainties remain when distinguishing between
infall, rotation (Keplerian or otherwise), and outflow.  Many of the
objects show velocity gradients that are consistent with Keplerian
rotation, but could also be attributed to another bulk motion, such as
infall. Furthermore even when rotation is indeed Keplerian, it is
difficult to distinguish the disk edge from its natal core
\citep{2005ApJ...628L.151D, Ces2005, 2006ApJ...642L..57D}. This, and
confusion with the outflow \citep{2006ApJ...642L..57D}, may cause disk
masses and extents to be overestimated.  The mass, luminosity and
multiplicity of the central object involve further uncertainties.

We have already noted that the angular momentum seen in observations
coincides with our estimated upper bound from the core collapse model.
Although we think this agreement is likely to be real, it is also
possible that the observed rotation is about a stellar group rather
than a single object.  This alternative is strengthened by the fact
that many rotating envelopes are inferred to be more massive than any
single central object, on the basis of the central luminosity.  Such
overweight disks are subject to strong {\em global} self-gravitational
instabilities quite distinct from the local instability we addressed
in \S \ref{Dfrag}.  We refer the reader to \cite{Shu1990} on this
point; see also the discussion in \S \ref{finitemass} and that given
in ML05.

It is notable that the three observations most confidently interpreted
as thin massive-star disks
\citep{Pat2005,2005ApJ...628L.151D,Shep2001} are the best match to our
predictions.  The other, more extended massive disks or toroids are of
uncertain mass and radius and may enclose many stars. As noted by
\cite{Ces2005} and \cite{Beuth}, in these instances we might be seeing
proto-cluster, rather than protostellar, disks and toroids.

\section{Fragmentation of Core-Collapse Disks} \label{coreval}

A numerical evaluation of fragmentation is presented below in \S
\ref{Results}, but first we calculate the scalings that govern these
results.  Given the fragmentation criterion from \S \ref{criteria} and
the fiducial core-collapse model described in \S \ref{coremod}, we can
estimate the ratio $T_d/T_{\rm crit}$ by ignoring either irradiation
(``active'' disks, as in \S \ref{visc}) or viscous heating
(``passive'' disks, using \S \ref{irrad}):
\begin{equation} \label{Scalings}
{T_d(R_d)\over T_{\rm crit}} = \left\{
\begin{array}{ll}
0.15 \left(\varepsilon_{0.5}^{7.7} \kappa_{R,\rm cgs}^4 \Sigma_{\rm
  cgs}^5  \over M_{\star f,30}^7\right)^{1/20}, & (\mbox{active
  disk})\\ 
0.35 \left(L_5\over \varepsilon_{0.5}^{1/60} M_{\star f,30}^3
  \Sigma_{\rm cgs}^{1/4} \right)^{1/4}, & (\mbox{passive disk})
\end{array}
\right.
\end{equation} 
where $L_\star = 10^5 L_5 L_\odot$, $\varepsilon=0.5
\varepsilon_{0.5}$, $\kappa_R = \kappa_{\rm R,cgs}$\,cm$^2$\,g$^{-1}$,
and $\Sigma=\Sigma_{\rm cgs}$\,g\,cm$^{-2}$. 

Note that increasing $M_{\star f}$ destabilizes disks in both regimes.
High values of $\Sigma$ enhance viscous heating relative to
irradiation, but neither is sufficient to prevent fragmentation around
a 30 $M_\odot$ star when $\Sigma\sim 1$\,g\,cm$^{-2}$.  However, if we
take $L=20$ and $\Sigma = 0.034$\,g\,cm$^{-2}$ -- values typical of
the low-mass star formation studied by ML05 -- then $T_d(R_d)>T_{\rm
crit}$ for $M>1.3\,M_{\odot}$ according to equation (\ref{Scalings}).
The evaluation used here neglects several effects treated in that
paper, such as thermal core support.  Nevertheless these scalings
explain why disks are intrinsically less stable during massive star
formation than in the low-mass case, as seen in detail below.

\subsection{Results} \label{Results} 

\begin{figure} 
\includegraphics[scale=.45]{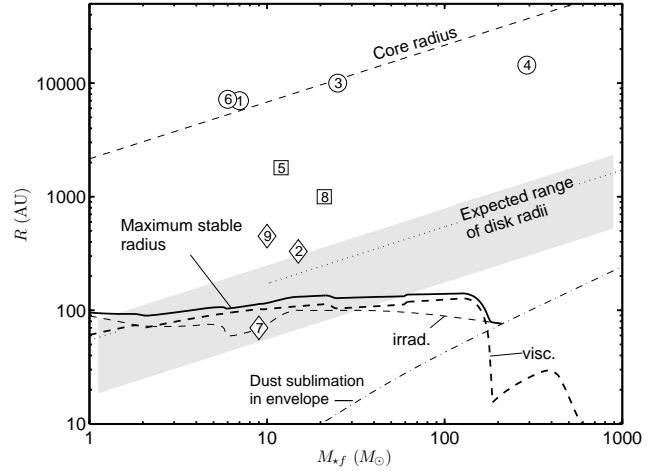}
\caption{Relevant radii. The characteristic disk radius predicted by
  (fiducial) core accretion is accompanied by a shaded band
  illustrating our (Maxwellian) model for its dispersion.  The largest
  stable radius is plotted for comparison; this is accompanied by the
  contributions from pure irradiation (no viscous heat) and pure
  viscosity (no irradiation).  The turnover of $R_{\rm crit}$ above
  $\sim 20\Msun$ is due to the scaling of ZAMS mass and
  luminosity. The references for the observational points are listed
  in table (\ref{reftab}). Circles represent objects that may best be
  described as cores, whereas diamonds represent those objects whose
  disks are well resolved. Squares indicate objects for which it is
  unclear whether they are rotating, infalling, or both; see \S
  \ref{observations} for discussion. }\label{frag}
\end{figure}

Fragmentation is expected when $R_{d,f}>R_{d,\rm crit}$, i.e., when
the disk extends past the critical fragmentation radius at some point
during formation.  Figure \ref{frag} compares these two radii for
a range of stellar masses, using the fiducial, conservative core
collapse model ($k_\rho=3/2$, $\varepsilon = 0.5$, $\Sigma_{\rm
cl}=1$\,g\,cm$^{-2}$, $\theta_j = 0.23f_j$) to compute $R_d$.

To address the lowest-mass stars whose disks can fragment, we must
account for the thermal component of a core's hydrostatic pressure and
for the accretion luminosity, both of which are negligible in very
massive stars.  We have (1) included a thermal component (at 20 K) in
the effective temperature that sets the accretion rate, so that
$\dot{M}_{\star d}\geq 10^{-5.3} \varepsilon
M_\odot$\,yr$^{-1}$ for all masses; (2) added accretion luminosity to
the ZAMS luminosity when estimating disk irradiation; and (3) employed
the \cite{1992ApJ...392..667P} models (for $\dot{M}_{\star d} =
10^{-4} M_\odot$ yr$^{-1}$, which is appropriate) to estimate the
radius of the accreting protostar.  Although rather approximate, these
amendments are of diminishing importance as $M_{\star f}$ increases
beyond $\sim 20M_\odot$.

Given the expected range of disk radii, all the disks presented in
figure \ref{frag} are candidates for fragmentation.  The expected disk
radius crosses the fragmentation boundary for $M_{\star f}\simeq 3.5
M_\odot$, and the two remain almost equal until $M_{\star f}\simeq 10
M_\odot$; fragmentation is marginal in this range.  Fragmentation
becomes increasingly likely as the mass increases, though slowly:
$R_{\rm crit}$ is within a factor of 2 of $R_d$ for $M_{\star f} < 23
M_\odot$.  For $M_{\star f}> 57M_\odot$, $R_{\rm crit}$ drops below
the range of disk radii implied by the dispersion of $f_j$ given below
equation (\ref{theta_j_evaluation}) -- as indicated by the gray region
in figure \ref{frag}.
The specific masses quoted depend on our model for angular
momentum, particularly as the critical radius is relatively constant
in the range 100-150 AU.

Recall, however, that the disk angular momentum derives from a
turbulent velocity field and is therefore quite stochastic.  The
spread in $j$ predicted by a Gaussian model for the velocity field
allows for the frequent formation of disks twice as large as predicted
in equation (\ref{Rd_from_Rc}).  Likewise, much smaller disks (by
about a factor of nine) can form equally easily from a chance
cancellation within the core velocity field.  This dispersion in
expected radii is indicated as a shaded band in figure \ref{frag}.
Remember also that we adopted a conservative estimate of the disk
angular momentum; otherwise, disk fragmentation would have been even
more prevalent.  Taking these points into account, we can draw a few
conclusions with relative certainty.
\begin{enumerate}%\begin{itemize}
\item \label{Os_fragment} A {\em significant portion} of the O star
  and early B star protostellar disks predicted in the core collapse
  model are prone to fragmentation, although the exact fraction is
  sensitive to the (uncertain) angular momentum scale and
  fragmentation criterion;
\item \label{Massives_Fragment} The tendency of disks to fragment
  increases with stellar mass ($M_{\star f}$); it decreases with
  higher column densities ($\Sigma_{\rm cl})$, and with steeper
  initial density profiles ($k_\rho$). 
\item \label{cutoff} Disks accreting more rapidly than about
  $1.7\times 10^{-3} \Msun\,$yr$^{-1}$ are destabilized by the sharp
  drop in dust opacity at $\sim 1050$ K, according to equation
  (\ref{cscrit}).  In the fiducial core model, this occurs above about
  110 $M_\odot$, close to the observed upper limit of stellar masses.
  More generally, this occurs for $M_{\star f}>87 (t_{\rm
  acc}/10^5{\rm yr}) \Msun$. 
\item \label{sublimation} At somewhat higher accretion
  rates, however, dust sublimation invalidates our model for starlight
  reprocessing in the infall envelope. 
\item \label{opacity_suppreses} So long as disks remain optically
  thick, any effect that decreases the Rosseland opacity is
  destabilizing.  For instance, low-metallicity disks are less stable
  than those of solar composition.  \citep[Primordial disks are
  however opaque in their inner portions.][]{2004ApJ...603..383T} 
  By adopting the (relatively
  opaque) \cite{2003A&A...410..611S} dust opacities, we have
  underestimated disk fragmentation. 
\end{enumerate} %\end{itemize}

\begin{table}
\begin{tabular}{|c|l|}
\hline
Number in figure \ref{frag} & Reference\\
\hline
1 & \cite{2005AA...434.1039C}\\
2 & \cite{Pat2005}\\
3 & \cite{2003AA...407..225O}\\
4 & \cite{2003AA...407..225O}\\
5 & \cite{2002ApJ...566..982Z}\\
6 & \cite{1999AA...350..197B}\\
7 & \cite{Shep2001}\\
8 & \cite{Shep2001}\\
9 & \cite{2005ApJ...628L.151D}\\
\hline
\end{tabular}
\caption{Observational data points in figure \ref{frag} and
corresponding references.} \label{reftab}
\end{table}

\subsection{Effect of Varying Efficiency}\label{CoreEffic}
%\input{CoreEffic.tex}
%\subsection{Accretion Rates vs Luminosity}\label{stromgren}
%\input{AccnVsL.tex} 
Up to this point we have adopted $\varepsilon= 0.5$ as the fiducial
accretion efficiency, following \cite{MT2003}.  In the theory of
\cite{MM2000}, $\varepsilon$ is set by the ejection of material by a
centrally-collimated protostellar wind. \citeauthor{MM2000} show that
$\varepsilon$ is quite insensitive to the ratio of infall and outflow
momentum fluxes.  Nevertheless, 0.5 is only an estimate and
$\varepsilon$ could well vary during accretion.   This is especially
true if the protostellar wind were ever to truncate accretion, as 
$\varepsilon(t)\rightarrow 0$ when this happens.  We briefly 
consider other values here. 

The primary effect of varying $\varepsilon$, while fixing $M_{\star
f}$, $\Sigma_{\rm cl}$, and $k_\rho$, is to change the core mass
required to make a star of that mass.  Suppose we halve $\varepsilon$,
so that $M_c$ must double.  The accretion time then increases, and
$\dot{M}_{\star d}$ decreases, by a factor $2^{1/4}$ (for
$k_\rho=3/2$).  This mildly stabilizes the disk.  But at the same
time, $R_c$ has been increased by $2^{1/2}$, $j$ has gone up by
$2^{3/4}$, and $R_{d,f}$ has expanded by $2^{3/2}$.  Balancing these
contributions, we expect lowering $\varepsilon$ to destabilize the
disk.   This was predicted also in equation (\ref{Scalings}), where
lowering $\varepsilon$ is seen to decrease stability in an active
disk.  Passive disks are extremely insensitive to $\varepsilon$. 

\begin{figure} 
\centering
\includegraphics[scale=.45]{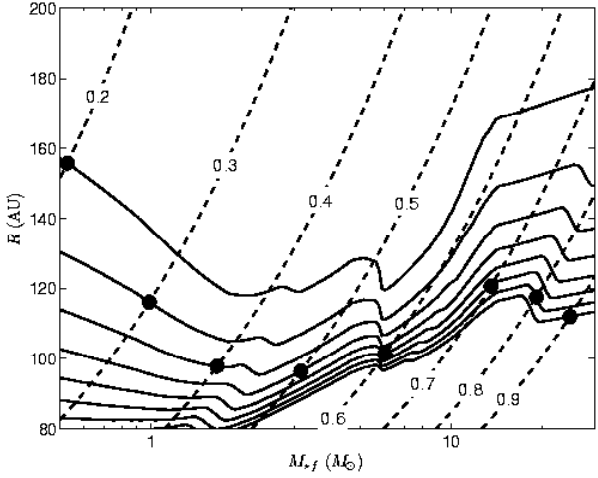}
\caption{Effect of varying the star formation efficiency $\varepsilon$
  in the fiducial core collapse model. {\em Dashed lines}: critical
  disk radius $R_{d,\rm crit}$ for fragmentation; {\em solid lines}:
  expected disk extent $R_d$. Intersections are as marked. } \label{diskseff} 
\end{figure}

Figure (\ref{diskseff}) corroborates our expectation by showing that
lower values of $\varepsilon$ correspond to less stable disks.
Indeed, the mass at which fragmentation sets in is 
sensitive to $\varepsilon$, specifically, $M_{\rm crit}
\propto\varepsilon^{2.6}$, while the critical disk radius is
relatively constant.  Does this mean that a decline in core efficiency
over time destabilizes disks?  Probably not, since most of the mass
will be accreted at an intermediate value of $\varepsilon$ (see the
discussion below equation \ref{rcoremass}).

\subsection{Time Evolution of Disk Fragmentation} \label{time}
For those disks that do suffer fragmentation, it is useful to know
whether this happens early or late in accretion and how much matter is
potentially affected.  To address these questions we construct the
time history of the accretion rate, stellar mass, and disk radius for
a star with $M_{\star f}=30\,\Msun$ in the fiducial core collapse
model.  For this calculation we use the luminosity history of such a
star as presented by \citet{MT2003}.  The scalings $R_d(t)\propto
M_\star(t)^{1/(3-k_\rho)}$ and $\dot{M}_{\star d}(t) \propto
M_\star(t)^{(6-2k_\rho)/(6-3k_\rho)}$ permit us to gauge disk
fragmentation through time.

\begin{figure} 
\centering
\includegraphics[scale=.45]{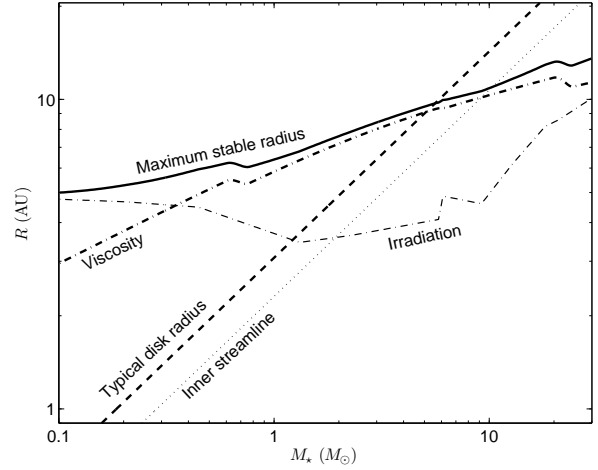}
\caption{Growth of disk and critical radius with the mass of a
  protostar accreting toward $30\Msun$ in the fiducial core collapse
  model.  In addition to the expected disk radius, we show the
  splashdown radius of the inner infall streamline, calculated
  assuming $\varepsilon=0.5$.  }\label{timev}
\end{figure}

Figure (\ref{timev}) shows the evolution of $R_d$ and $R_{\rm crit}$
during accretion.  In this plot, the relative constancy of the
fragmentation radius is due to the enhanced effect of irradiation at
low masses.  We find that a star destined to become $30\Msun$ (born of
a 60 $M_\odot$ core) has a disk that crosses into the regime of
fragmentation when the protostar has accreted approximately $5.6\Msun$.
The fact that this is slightly higher than the critical mass
identified earlier is to be expected: the inner $5.6\Msun$ of a
larger core is equivalent to a $5.6 \Msun$ core with a slightly higher
column density -- specifically, $\Sigma_c\propto (M_{\star}/M_{\star
f})^{-1/3}$, for $k_\rho = 3/2$ (cf. equation~\ref{Sigma_sph(m)}).
The somewhat higher column density implies a smaller and somewhat
stabler disk, leading to a slightly higher mass scale for fragmentation. 

We also show on figure (\ref{timev}) the radius of the innermost
streamline of infalling material from the envelope from
\cite{1984ApJ...286..529T}, again assuming an accretion efficiency of
$50\%$.  This, along with stochastic variations in disk angular
momentum about its typical value, suggest that accretion can coexist
with fragmentation so long as the disk is not too far beyond the
fragmentation threshold; see \S \ref{diskeff} for more discussion.

\subsection{Disk mass and global instability} \label{finitemass} 

At the typical fragmentation radius of $100-150$ AU, the mass scale of
a disk with $Q=1$ is 
\begin{equation} \label{massscale}
{\pi R_d^2 \Sigma_d(R_d) \over M_{\star f}}= 0.10 \varepsilon_{0.5}^{1/12}
  \left(R_d\over 150\,\rm AU\right)^{1/2} {\Sigma_{\rm cgs}^{1/4} \over
  M_{\star f,30}^{1/4}}.
\end{equation} 
Global instabilities of the disk are triggered by the total disk mass
\citep{1989ApJ...347..959A,Shu1990}, which is larger than $\pi R_d^2
\Sigma_d(R_d)$ by the factor $2/(2-k_\Sigma)$ if $\Sigma_d\propto
r^{-k_\Sigma}$ within $R_d$.  There is therefore the possibility that
the fast angular momentum transport by these modes
\citep{1998ApJ...504..945L} suppresses local
fragmentation, but we consider it unlikely that fragmentation is
eliminated by this process.

\section{Consequences of Instability}\label{fragments}

\subsection{Fragment Masses} \label{fragmass}

Once a disk fragments, what objects form?  \cite{2004ApJ...608..108G}
determine the {\em initial} fragment mass based on the wavenumber of
the most unstable mode in the disk. The corresponding wavelength of
this axisymmetric mode is:
\begin{equation}
\lambda(r)=2\pi\frac{c_{\rm ad}^2}{\pi G \Sigma},
\end{equation}
where $\Sigma$ and $c_s$ are both functions of radius within the disk.
Assuming that the fragment has comparable dimension azimuthally, the
corresponding mass scale is \citep{2004ApJ...608..108G} 
\begin{eqnarray} \label{mfrag}
M_{\rm frag}& =& \lambda^2\Sigma
\nonumber \\ 
&=&  \frac{4 \pi}{\Omega} \frac{c_{\rm ad}^3}{G} Q, 
\end{eqnarray}
i.e., roughly the amount of mass accreted in $2Q$ orbits.  We assume
fragmentation only occurs when $Q\rightarrow 1$.  We consider our
rather idealized estimate of $M_{\rm frag}$ uncertain by at least a
factor of two.  In reality the initial mass is likely a stochastic
variable, best determined from numerical simulations (R. Rafikov,
private communication).

Once a fragment forms, its growth is controlled by accretion of
surrounding gas, migration through the disk, and collisional or
gravitational interaction with other fragments.  Rather than address
these questions in detail, we will draw preliminary conclusions by
comparing $M_{\rm frag}$ to two critical scales: the gap opening mass
$M_{\rm gap}$ and the isolation mass $M_{\rm iso}$.  A fundamental
uncertainty is the state of the gas disk: again, we assume $Q=1$.

When the gravitational torques exerted on the disk by the fragment
exceed viscous torques, a gap opens around the fragment.
\cite{2002ApJ...572..566R} estimates
\begin{eqnarray} \label{mgap}
M_{\rm gap}&=& \frac{2 c_{\rm ad}^3}{3 \Omega G}\frac{\alpha}{0.043} 
\nonumber \\ 
&=& 0.37 \frac{\alpha/0.30}{Q} M_{\rm frag}. 
\end{eqnarray}
Since this is less than $M_{\rm frag}$, and gets even smaller if the
disk viscosity goes down and thus cooling time goes up relative to the
critical state in the \cite{Gam2001} simulations, we expect fragments
to open gaps immediately.

Gap opening slows but does not necessarily end accretion
\citep{1996ApJ...467L..77A}.  A possible limit to growth is set by the
point at which the fragment accretes all the mass within its Hill
radius \citep[e.g.,][but see
\citeauthor{1996ApJ...467L..77A}]{2004ApJ...608..108G}.  This defines
the isolation mass
\begin{equation} \label{miso}
M_{\rm iso}(r)\approx \frac{(2\pi f_H r^2 \Sigma)^{3/2}}{9 M_\star^{1/2}}
\end{equation}
where $f_H\simeq 3.5$. 

\begin{figure} 
\includegraphics[scale=.45]{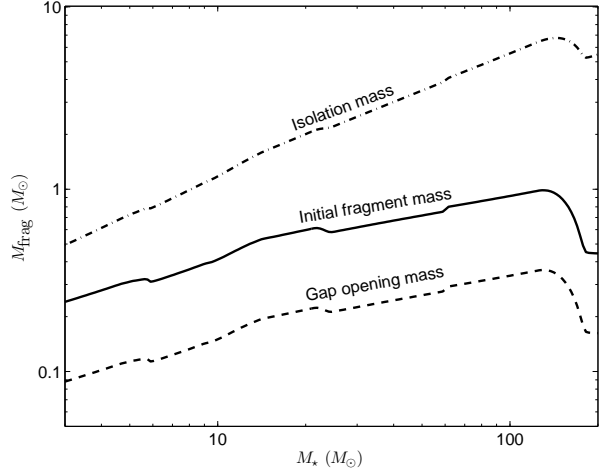}
\caption{Our estimate of the initial fragment mass, compared to the
  gap opening mass and isolation mass, at the end of accretion in the
  fiducial core collapse model. We truncate the calculation where dust
  sublimation in the envelope makes the critical radius determination
  uncertain.}\label{fragrat}
\end{figure}

Figure (\ref{fragrat}) compares the initial, gap opening, and
isolation masses for a range of $M_{\star f}$, evaluated at $R_{\rm
crit}$ near the end of core accretion (in the fiducial core model).
Because gap opening slows accretion, we expect the masses of disk-born
stars to resemble $M_{\rm frag}$ more than $M_{\rm iso}$.
In this case they will be low-mass stars of order 0.2--0.4 $\Msun$.

Fragmentation tends to set in at $\sim 100-200$ AU, as
we noted in \S \ref{nomodel}.  However gap-opening fragments should be
swept inward if disk accretion continues; only a single disk mass of
accretion is required to bring them in to the central object.  If they
move inward by about a factor of 30, they will be within a few stellar
radii at carbon ignition \citep{1985ibs..book...39W}.  As candidates
for mass transfer and common-envelope evolution, such objects can
serve as reservoirs of matter and angular momentum for the relic of
the central star's supernova explosion.

\subsection{Observability of Disk-Born Stars}\label{observations}

We have predicted that stars with extended protostellar disks will
produce low mass (M5--G5) companions.  Thus we would expect many O
stars, and perhaps some early B stars, to have multiple coplanar
companions at separations of order $\ltsim 100-200$ AU, depending
on the amount of migration that occurs following formation. In the
closest clusters, for example, Orion, this correspondes to an angular
separation of approximately $0.4''$ and an apparent bolometric
magnitude difference of, at a minimum, $\sim 13$ magnitudes. This
implies that even in K-band with an AO system such as that of VLT,
such objects would be difficult to observe (M. Ahmic, private
communication). Similarly, the combination of AO with a coronograph
(e.g. the Lyot coronograph on AEOS) can provide a dynamic range of up
to up to 8 H-band magnitudes at a few hunderd mas, which is still too
small to detect the aforementioned companions
\citep{2006astro.ph..9337N}.If not detectable during the main sequence 
life of the primary star, the presence of such companions might be
observable via binary interaction once the primary evolves.

\subsection{Disk Efficiency}\label{diskeff}

If a $30\Msun$ star will suffer disk fragmentation early in 
its accretion, as we estimated in \S \ref{time} on the basis
of the turbulent core model, then we must address how this might
impact subsequent accretion.  One possibility is that gap-opening
fragments will be swept inward to the central star, in which case its
final mass will be unaffected.  This outcome resembles the scenario
outlined by \cite{2003astro.ph..7084L} for gravitationally unstable
AGN accretion.  Alternatively, \cite{2005MNRAS.362..983T} suggest
low-luminosity AGN may be starved of gas by fragmentation.

In the latter scenario, a strongly unstable disk will have a low {\em
disk efficiency}, $\varepsilon_d \equiv \dot M_\star/\dot M_{\star
d}$.  We can estimate $\varepsilon_d$ in the limit that none of the
gas entering an unstable region of the disk ultimately accretes onto
the star.  The splashdown radius of the innermost streamline was
estimated in expression (\ref{Rinner}). Given that outflow removes
matter from the inner streamlines, and that fragmentation removes it
from the outer portions of the disk, the fraction of mass that
successfully accretes (after striking the disk) is 
\begin{equation}\label{epsilon_d} 
 \varepsilon_d 
%% = \frac{\left(1-\frac{R_{\rm
%%  crit}}{R_d}\right)^{1/2}+\varepsilon-1}{\varepsilon}
 = 1- \varepsilon^{-1}\left[1- \left(1-\frac{R_{\rm
 crit}}{R_d}\right)^{1/2}\right]. 
\end{equation} 
Of course, this expression is only applicable when it
yields $0\leq \varepsilon_d\leq 1$, i.e., when the disk is partly but
not wholly unstable.

Two complications arise when evaluating equation (\ref{epsilon_d}).
First, the critical radius $R_{\rm crit}$ must be calculated using the
mass accretion rate outside of itself.  Second, recall that the
angular momentum of infalling gas derives from its initial turbulent
velocity and is likely to vary in direction and magnitude.  The
fluctuations of $j$ were estimated by the distribution $f_j$, which
appeared in equation (\ref{theta_j_evaluation}).  Accounting for both
of these effects, we find
that accretion can continue even in an actively unstable disk. For
$30\Msun$, we find that the infall streamline remains within the
stable disk radius through the end of accretion. Even if all the gas
entering a fragmenting region is consumed, this need not fully starve
the central object.  However, the tendency to fragment becomes much
stronger for 
more massive stars.  This is especially true for those ($M_{\star
f}\sim 110 \Msun$, from \S \ref{coreval}) that accrete rapidly enough
that their disks are destabilized by the drop in dust opacity.
%
%%Although the results quoted above assume a 50\% core formation
%%efficiency, we have investigated the effect
%%various core efficiencies on disk efficiency. Low efficiencies produce larger
%%expected disk radii,and thus larger radii for the innermost
%%streamlines. However, both very high and very low ($\geq 60\% 
%%and \leq 30\%$) core efficiences lead to larger critical radii. As a result
%%disk efficiences are slightly higher. Correspondingly, core
%%efficiencies close to $50\%$ produce lower disk
%%efficiencies.

In reality, we expect some gas to accrete through the unstable
region.  A rough upper limit would be to adopt the accretion
rate for a Toomre-critical ($Q=1$) state, and assume that any surplus
is consumed by fragmentation.  In this case, the central accretion
rate would be limited by the temperature of the coldest region of the
disk (according to eq.~\ref{cscrit}).  Numerical simulations will
ultimately be required to quantify the behaviour of Toomre-unstable
disks. 

Furthermore, there are numerous feedback mechanisms that have not been
addressed. For example, if fragmentation halts accretion, this could
change the outflow power, the shape of the outflow and infall cavity
and thus the heating of the disk by reprocessed starlight. This
interplay will be investigated in future work.

\section{Discussion} \label{discussion}
Our primary conclusion (\S \ref{nomodel}) is that massive-protostar
disks that accrete more slowly than $\sim 1.7\times 10^{-3}
\Msun$~yr$^{-1}$ are subject to fragmentation at disk radii beyond
about 150 AU.  This critical radius is set primarily by the viscous
heating of the disk midplane as it accretes, with reprocessed
starlight playing an equal or secondary role for most stellar masses.
As all the disks we consider are optically thick, the critical radius
depends on the Rosseland opacity law $\kappa_R(T)$ within dusty disk
gas.

Comparing to our conservative estimate of the disk radius in the
\cite{MT2003} model for massive star formation by the collapse of
turbulent cores (\S \ref{coremod}), we find that fragmentation is
marginal for stars accreting four to 15 solar masses; higher-mass
stars are increasingly afflicted by disk fragmentation.  Although the
mass at which fragmentation sets in is sensitive to our somewhat
uncertain fragmentation criterion (\S \ref{criteria}) and angular
momentum calculation (\S \ref{angmom} and the Appendix), we have been
conservative in five ways: (1) by adopting a fragmentation temperature
lower than that implied by the \cite{2005MNRAS.364L..56R} simulations;
(2) by adopting a low estimate of the specific angular momentum that
determines the disk radius; (3) by adopting a relatively opaque model
for the disk's Rosseland opacity; (4) by ignoring the shielding effect
of a moderately opaque infall envelope, and (5) by adopting a low
estimate for the cooling time, and a hard equation of state.  All of
these approximations should, if anything, lead us to underestimate the
prevalence of disk fragmentation.  Along with the existence of
turbulent fluctuations in $j$ (\S \ref{coreval}), these points ensure
that {\em some} massive stars above $\sim 10\Msun$ experience disk
fragmentation.   As noted in \S
\ref{finitemass}, we cannot rule out the possibility that global
instabilities flush disk material fast enough to suppress
fragmentation, but we consider it unlikely that this prevents {\em
  all} fragmentation. 

 We therefore expect multiple, coplanar, low-mass (M5 to G5, \S
\ref{fragmass}) companions to form around many O (and some B) stars.
Given initial separations of order 100-200 AU, their photospheric
emission is not observable with present techniques.  Disk migration,
followed by mass transfer or common-envelope evolution, may however
make them evident as the primary evolves (\S \ref{observations}).

Even when disks fragment, we expect some accretion onto the central
star -- if only because of material that falls within the
fragmentation radius (\S \ref{diskeff}).  Although we cannot yet
quantify the disk efficiency parameter $\varepsilon_d$, we expect it
to be significantly less than unity for those early O stars ($M_{\star
  f}\gtsim 50 \Msun$) whose disks are most prone to fragment. 

%% If massive stars accrete via competitive Bondi-Hoyle accretion, our
%% argument at the beginning of \S \ref{core_vs_comp} implies their disks
%% will probably be too small to fragment.  Observations of high angular
%% momentum (\S \ref{diskobs}) support the turbulent core model, but this
%% depends on the observed gas being bound to a single object or binary. 

\subsection{Imprint on the initial mass function} \label{IMFlimit} 

The stabilizing effect of viscous heating is absent in disks that
accrete more rapidly than $1.7\times 10^{-3}\Msun$\,yr$^{-1}$, thanks
to a sharp drop in dust opacity at $\sim 1050$ K (see \S
\ref{coreval}).  As this affects stars of mass greater than 87$(t_{\rm
acc}/10^5$yr$)\Msun$, or about 110 $\Msun$ in the fiducial core model,
it may be related to the cutoff of the initial mass function (IMF) at about
120 $\Msun$. 

Several other explanations for this cutoff have been proposed,
all involving the increasing bolometric luminosity, ionizing
luminosity, or outflow force emitted by the central star.  Starvation
by disk fragmentation has the distinctive feature that it becomes much
more severe at a specific accretion rate.  For this reason we expect
it to produce a sharper IMF cutoff.  (The transition to super-Eddington
luminosities could also produce a sharp cutoff, but
\citealt{2005ApJ...618L..33K} argue that this can be overcome by
asymmetric radiation transfer.) 

Note also that disk accretion is {\em destabilized} by rapid
accretion, whereas rapid accretion quenches the effects of direct
photon force and of the ionizing radiation \citep{WC1987}.  Disk
fragmentation may close an avenue by which very massive stars would
otherwise form.

\subsection{Proto-binary disks}
\cite{PS2006} state that close massive binaries $< 10 $ year periods)
are more likely to have nearly equal masses. More generally, the
binarity fraction overall among massive stars is higher than their low
mass counterparts ($1.5 \mbox{ versus } 0.5$,
\cite{2005IAUS..227...12B}). Due to the high fraction of roughly equal
mass binaires, their effect on disk dynamics must be addressed in
future work. Moreover, the stellar densities in regions of HMSF are
high, suggesting that other cluster stars might be close enough to
interfere with disks stretching out to $\sim$ 100 AU
\citep{2003MNRAS.339..577B}.  It is not currently clear whether disk
accretion can preferentially grow a low-mass companion until its mass
rivals that of the primary star, as \cite{1996ApJ...467L..77A}
suggest.  If so, then disk fragmentation may be relevant in the
production of equal-mass binaries; if not, they must form by another
mechanism.  In any case, the multiplicity of the center of gravity
should be accounted for in future work.  It seems unlikely, though,
that a binary with $\sim 10$ year period will stabilize the
fragmenting regions whose periods are $\gtsim 300$ years.

\noindent {\bf Acknowledgments:} We are pleased to thank the referee,
Jonathan Tan, for very useful suggestions, as well as Ray
Jayawardhana, Scott Kenyon, Chris McKee, Stefan Mochnacki, Roman
Rafikov, Debra Shepherd, and Yanqin Wu for helpful discussions, and
Alyssa Goodman for a leading question. We would also like to thank
Susana Lizano for insightful suggestions regarding disk
observations. C.D.M.'s research is funded by NSERC and the Canada
Research Chairs program. K.M.K.\ is supported by a U. Toronto
fellowship.

\onecolumn
\section{Appendix} \label{appendix}

\newcommand{\rvec}{{\mathbf r}}
\newcommand{\vvec}{{\mathbf v}}

We present here an estimate of the angular momentum of cores initially
supported by turbulent motions, generalizing the results of ML05 to
arbitrary line width-size relations.  The analytical treatment of this
problem rests on several idealizations about turbulent velocities: (1)
that they are isotropic and homogeneous; and (2) that the Cartesian
velocity components are neither correlated with each other, nor (to
any appreciable degree) with density fluctuations.  For simplicity we
also assume (3) that the core density profile is spherically symmetric
and can be captured in a single function $\rho(r)$.  When evaluating
our formulae we assume (4) that the velocity difference between two
points scales as a power of their separation, and (5), for the purpose
of computing fluctuations, that the velocity components
are Gaussian random fields.

One might object that condition (1) is inconsistent with the turbulent
support of an inhomogeneous density profile.  Consider, however, that
if the core profile is a power law $\rho(r)\propto r^{-k_\rho}$, then
the turbulent line width must scale as $\sigma(r)\propto r^\beta$
where 
\begin{equation}\label{beta}
\beta = 1-k_\rho/2. 
\end{equation}
A velocity field with this scaling is consistent with our conditions
(1), (2), and (4) if 
\begin{equation}\label{dv-dr}
\langle [\vvec_i(\rvec_1)-\vvec_j(\rvec_2) ]^2 \rangle = k
|\rvec_1-\rvec_2|^{2\beta} \delta_{ij}
\end{equation} 
where angle brackets represent an ensemble average and $k$ is a
normalization constant related to the virial parameter $\alpha$.  ML05
considered the case $\beta=1/2$ appropriate for giant
molecular clouds ($k_\rho=1$);  we generalize their formulae to other
values of $\beta$, including the fiducial value $\beta=1/4$ 
corresponding to $k_\rho=3/2$. 

Our goal is to compute the expectation value and dispersion of the
core specific angular momentum $j$, normalized to $R_c \sigma$ where
$\sigma$ is the one-dimensional line width of the core.  We find
that, under our assumptions, 
\begin{equation} \label{<j^2>}
\langle j^2 \rangle = -\int \int \frac{d^3 \rvec_1 \rho(r_1)}{M}
\frac{d^3 \rvec_2 \rho(r_2)}{M} \rvec_1\cdot\rvec_2 
 \langle[v_x(\rvec_1)-v_x(\rvec_2)]^2\rangle
\end{equation} 
and that 
\begin{equation} \label{<sigma^2>} 
\langle \sigma^2 \rangle = \frac{1}{2}\int \int \frac{d^3 \rvec_1 \rho(r_1)}{M}
\frac{d^3 \rvec_2 \rho(r_2)}{M}
\langle[v_x(\rvec_1)-v_x(\rvec_2)]^2\rangle 
\end{equation} 
where $M=\int \rho d^3 \rvec$ and all integrals are restricted to the
region of interest (typically the core interior).  These equations,
which we will prove below, agree with the formulae in ML05's Appendix
but allow $\langle[v_x(\rvec_1)-v_x(\rvec_2)]^2\rangle$ to be an
arbitrary function of $|\rvec_1-\rvec_2|$.  It is important to note
that the two formulae are identical up to the factor $-2
\rvec_1\cdot\rvec_2$ in the integrand. The negative sign in equation
(\ref{<j^2>}) ensures that $\langle j^2\rangle$ is positive, since
$\langle[v_x(\rvec_1)-v_x(\rvec_2)]^2\rangle$ takes higher values when
$|\rvec_1-\rvec_2|$ is large, hence when $\rvec_1\cdot\rvec_2$ is
negative.

\begin{figure} \label{jForCores}
\centering
\includegraphics[scale=.45]{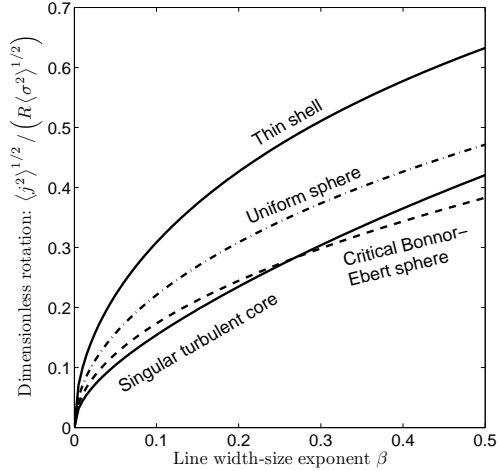} %{JvsSigmaR.eps}
\caption{Values of $\langle j^2\rangle^{1/2}/(R\langle
  \sigma^2\rangle^{1/2})$ evaluated for turbulence with line
  width-size exponent $\beta$ and three relevant density profiles.
  For hydrostatic turbulent cores, $\rho \propto r^{-2(1-\beta)}$.  }
\end{figure}

To evaluate equations (\ref{<j^2>}) and (\ref{<sigma^2>}) we make use
of spherical symmetry and impose equation (\ref{dv-dr}) for the
velocity correlations.  Defining $\mu = \rvec_1\cdot\rvec_2/(r_1r_2)$
as the cosine of the angle between $\rvec_1$ and $\rvec_2$, 
\begin{eqnarray} \label{<j^2>-angular} 
\langle j^2 \rangle &=& - k
\int_0^R dr_1 4\pi r_1^2
\frac{\rho(r_1)}{M}
\int_0^R dr_2 2\pi r_2^2 \frac{\rho(r_2)}{M}
r_1 r_2 
\left(r_1^2 + r_2^2\right)^{\beta} 
\int_{-1}^1 d\mu 
\left(1-q\mu\right)^{\beta} \mu
 \nonumber\\ &=& 
\frac{2 \pi^2 k}{M^2}\int_0^R dr_1 \int_0^R dr_2 
\left(r_1^2 + r_2^2\right)^{2+\beta} r_1 r_2 \rho(r_1)  \rho(r_2)
\left[\frac{(1+q)^{\beta+2}-(1-q)^{\beta+2}}{\beta+2}-
  \frac{(1+q)^{\beta+1}-(1-q)^{\beta+1}}{\beta+1} \right]
\end{eqnarray} 
and 
\begin{eqnarray} \label{<sigma^2>-angular} 
\langle \sigma^2 \rangle &=& \frac12 k
\int_0^R dr_1 4\pi r_1^2 \frac{\rho(r_1)}{M}
\int_0^R dr_2 2\pi r_2^2 \frac{\rho(r_2)}{M}
\left(r_1^2 + r_2^2\right)^{\beta}
\int_{-1}^1  d\mu \left(1-q\mu\right)^{\beta}
\nonumber\\
&=& \frac{2\pi^2 k}{M^2} \int_0^R dr_1 \int_0^R dr_2 
\left(r_1^2 + r_2^2\right)^{1+\beta} r_1 r_2 
\rho(r_1) \rho(r_2)
\left[\frac{ (1+q)^{\beta+1}-(1-q)^{\beta+1}}{\beta+1}\right]
\end{eqnarray} 
where $q = 2 r_1 r_2/(r_1^2+r_2^2)$, as in ML05.  
Table \ref{jOnRSigma} and figure \ref{jForCores} provide values of
$\langle j^2\rangle^{1/2}/(R\langle \sigma^2\rangle^{1/2})$ for
several density profiles, both for the internal spectrum given by
equation (\ref{beta}), and for $\beta=1/2$, which may better represent
the background spectrum.

\begin{table}\label{jOnRSigma}
\begin{tabular}{|l|l|l|l|l|l|}
\hline
Profile & $\beta = 1/4$ & $\beta=1/2$ \\
\hline
turbulent core &0.2704  &0.4206 \\ 
critical Bonnor-Ebert sphere &0.2730  & 0.3828 \\ 
uniform region ($k_\rho=0$) &  0.3430 & 0.4714 \\ 
thin shell & 0.4714 & 0.6324 \\ 
\hline
\end{tabular}
\caption{Values of $\langle j^2\rangle^{1/2}/(R\langle
  \sigma^2\rangle^{1/2})$ in our model for turbulent angular momentum. }
\end{table}

Figure \ref{jForCores} plots the ratio $\langle j^2\rangle^{1/2}
/(R\langle \sigma^2\rangle^{1/2})$ as given by equations
(\ref{<j^2>-angular}) and (\ref{<sigma^2>-angular}) for a turbulent
core profile, a critical Bonnor-Ebert sphere, and a thin shell.  The
results are very close to power laws in $\beta$: 
\begin{equation} \label{powerlaw-j-with-beta}
\frac{\langle j^2\rangle^{1/2}}{R\langle\sigma^2\rangle^{1/2}} \simeq
\left\{\begin{array}{ll} 0.655\beta^{0.638} & {\rm singular~turbulent~core} \\ 
0.537 \beta^{0.488} & {\rm critical~Bonnor-Ebert~sphere} \\ 
0.648\beta^{0.459} & {\rm uniform-density~sphere} \\ 
0.849\beta^{0.424} & {\rm thin~shell}.
\end{array}\right.
\end{equation}
The spin parameter $\theta_j$ makes reference to the velocity
dispersion $\sigma(R)$ at radius $R$, rather than the mean velocity
dispersion $\langle \sigma^2 \rangle^{1/2}$ within $R$.  For this we
use the scaling $\sigma^2\propto G M/r \propto
M^{2\beta/(3-k_\rho)}$ to derive 
\[ \frac{\sigma(R)^2}{\langle\sigma^2\rangle} =
1 + \frac{2\beta}{3-k_\rho}. \]
With this correction factor accounted for, $\theta_j/f_j$ is still
very close to a power law of $\beta$: 
\begin{equation} \label{powerlaw-thetaj-with-beta}
\frac{\theta_j}{f_j}\simeq
\left\{\begin{array}{ll} 
0.504\beta^{0.552} & {\rm singular~turbulent~core} \\ 
0.987 \beta^{0.651} & {\rm uniform-density~region}\\
0.849\beta^{0.424} & {\rm thin~shell}.
\end{array}\right.
\end{equation}
(The last line is unchanged, as there is no correction for a thin
shell.)  These formulae were quoted in \S \ref{angmom}. 

We now return to the derivation of equation (\ref{<j^2>}); equation
(\ref{<sigma^2>}) is treated below.  The $z$ component of the specific
angular momentum of the core is 
\begin{eqnarray} \label{j^2-initform} 
j_z^2 &=& \int d^3 \rvec_1 \int d^3 \rvec_2 \frac{\rho_1}{M} 
 (x_1 v_{y1}-y_1 v_{x1}) \frac{\rho_2}{M}  (x_2 v_{y2}-y_2 v_{x2})
\nonumber\\
&=& \frac{1}{M^2} \int d^3 \rvec_1 \int d^3 \rvec_2 \rho_1\rho_2 
(x_1 x_2 v_{y1} v_{y2} + y_1 y_2 v_{x1} v_{x2} - x_1 y_2 v_{y1} v_{x2}
- y_1 x_2 v_{x1} v_{y2}); 
\end{eqnarray} 
here, as below, we use subscripts 1 and 2 to indicate functions of
$\rvec_1$ and $\rvec_2$, i.e., $\rho_1 = \rho(\rvec_1)$.  On ensemble
averaging of the second line, the last two terms in equation
(\ref{j^2-initform}) become zero thanks to the $\delta_{ij}$ in
equation (\ref{dv-dr}).  The first two terms are equal, thanks to
spherical symmetry; thus 
\begin{equation} 
\langle j_z^2 \rangle = \frac{2}{M^2} \int d^3 \rvec_1 \int d^3
\rvec_2 \rho_1 \rho_2 x_1 x_2 \langle v_{y1} v_{y2} \rangle. 
\end{equation} 
Spherical symmetry allows us to replace $x_1 x_2$ with
$\rvec_1\cdot\rvec_2/3$ and $j_z^2$ with $j^2/3$, so 
\begin{eqnarray} 
\langle j^2 \rangle &=& \frac{2}{M^2} \int d^3 \rvec_1 \int d^3
\rvec_2 \rho_1 \rho_2 \rvec_1\cdot\rvec_2 \langle v_{y1} v_{y2}
\rangle \nonumber\\
 &=& \frac{1}{M^2} \int d^3 \rvec_1 \int d^3
\rvec_2 \rho_1 \rho_2 \rvec_1\cdot\rvec_2 
 \left( \langle v_{y1}^2 \rangle +\langle v_{y2}^2 \rangle  - \langle
 (v_{y1} - v_{y2})^2 \rangle \right). 
\end{eqnarray} 
The first two terms within the brackets are zero, as spherical
symmetry requires them to be even in the space coordinates while
$\rvec_1\cdot \rvec_2$ is odd; the surviving term yields equation
(\ref{<j^2>}). 

To prove equation (\ref{<sigma^2>}) we write 
\begin{equation} \label{sigma^2-initform} 
{M} \sigma^2 = \int d^3 \rvec {\rho(\rvec)}(v_x-\bar v_x)^2  
= \int d^3 \rvec \rho(\rvec) (v_x^2-\bar v_x^2)  
\end{equation} 
where $\bar \vvec = \int \vvec \rho(\rvec)/M d^3\rvec$ is the center
of mass velocity.  In equation
(\ref{j^2-initform}) we replace $\bar v_x^2$ with $\int \int
\vvec_1\cdot\vvec_2 (\rho_1/M)(\rho_2/M) d^3\rvec_1 d^3\rvec_2$, and
we bring the first term into similar form by multiplying it by $\int
(\rho_2/M)d^3\rvec_2 = 1$: 
\begin{eqnarray}
M \sigma^2 &=& \int d^3 \rvec_1 \int d^3 \rvec_2 \rho_1\rho_2 v_{x1}
(v_{x1}-v_{x2})
\nonumber \\ &=& \frac{1}{2} \int d^3 \rvec_1 \int d^3 \rvec_2 \rho_1\rho_2 
(v_{x1}-v_{x2})^2. 
\end{eqnarray} 
The second line follows from the first by noting that the integral is
antisymmetric under interchange of $\rvec_1$ and $\rvec_2$. When
averaged, it yields equation (\ref{<sigma^2>}).

\end{document}